# Experimental Observation of Hidden Multistability in Nonlinear Systems

Kun Zhang,[†] Qicheng Zhang,[†] Shuaishuai Tong, Wenquan Wu, Xiling Feng, and Chunyin Qiu[*]

Key Laboratory of Artificial Micro- and Nano-Structures of Ministry of Education
and School of Physics and Technology, Wuhan University, Wuhan 430072, China

[†]These authors contributed equally: Kun Zhang, Qicheng Zhang
[*]To whom correspondence should be addressed: cyqiu@whu.edu.cn

***Abstract.*** Multistability—the coexistence of multiple stable states—is a cornerstone of nonlinear dynamical systems, governing their equilibrium, tunability, and emergent complexity. Recently, the concept of hidden multistability, where certain stable states evade detection via conventional continuous parameter sweeping, has garnered increasing attention due to its elusive nature and promising applications. In this Letter, we present the first experimental observation of hidden multistability using a programmable acoustic coupled-cavity platform that integrates competing self-focusing and self-defocusing Kerr nonlinearities. Beyond established bistability, we demonstrate semi- and fully-hidden tristabilities by precisely programming system parameters. Crucially, the hidden stable states, typically inaccessible via the traditional protocol, are unambiguously revealed and dynamically controlled through pulsed excitation, enabling flexible transitions between distinct types of stable states. These experimental findings not only offer new insights into the fundamental physics of emerging hidden multistability, but also unlock new avenues for applications in information storage, information encryption, and safety precaution, where multi-state dynamics could enable advanced control techniques.

*Introduction.* Nonlinearity plays a crucial role in shaping the complex behaviors of dynamical systems, giving rise to intricate phenomena such as chaos, solitons and multistability. These phenomena, in turn, advance cutting-edge technologies and enhance fundamental understanding of natural processes [1]. In particular, the multistability refers to the coexistence of multiple stable states under a given set of parameters, with their emergence dependent on initial conditions. It has been observed across diverse scientific fields, including physics [2-10], chemistry [11,12], biology [13], neuroscience [14], ecology [15], genetics [16], and even climate science [17]. Typically, physical systems with multistability exhibit hysteresis under continuous parameter sweeping, and undergo transitions between different stable states due to random perturbations or deterministic controls. These properties make nonlinear systems particularly attractive for applications in information storage [18-21], switches [22-24] and logic gates [25-27].

Very recently, a fundamentally new class of multistability—hidden multistability—has been proposed theoretically [9]. Unlike the conventional multistability, it refers to the emergence of additional stable states that are folded within traditional hysteresis loops. As such, adiabatic parameter sweeping only guides the system along trajectories connecting explicit stable states [5], while leaving the embedded states hidden and unobservable. The existence of such hidden states could be a double-edged sword. On one hand, devices relying on known multistability may malfunction in real-world applications if random noise kicks the system into the hidden states [28-30]. On the other hand, such hidden states can securely store sensitive information, which makes unauthorized decryption exceedingly difficult without detailed system knowledge [31]. Therefore, identifying hidden multistability is crucial for understanding and regulating complex nonlinear systems. However, the experimental detection of hidden states, which are typically embedded in the system's hysteresis landscape, demands precise nonlinear control and unique excitation strategy, which pose substantial challenges for current experimental platforms and technologies.

In this Letter, we present the first experimental evidence of hidden multistability using a programmable nonlinear acoustic platform. We start with a minimal yet representative driven-dissipative double-oscillator model that features both self-defocusing and self-focusing Kerr nonlinearities. By carefully balancing these nonlinearities, our model exhibits not only conventional bistability but also intricate semi- and fully-hidden tristabilities at different driving frequencies. To experimentally validate these phenomena—especially the elusive hidden multistabilities—we design and implement a coupled acoustic binary-cavity platform integrated with external circuits. In this setup, the long-sought acoustic Kerr nonlinearities, characterized by intensity-dependent blueshifts and redshifts of resonant frequencies, are elegantly realized via active and programmable electroacoustic feedbacks—a key design to achieve precise nonlinear control. Under standard sound field sweeping, we observe a bistability-like hysteresis loop in the fully-hidden tristability system—while the hidden states remain concealed and undetectable. Crucially, by employing an unusual pulsed excitation protocol, we successfully reveal the elusive hidden states and demonstrate controlled transitions among multiple stable states. Our findings, aligned well with those predicted from nonlinear coupled-mode theory, not only validate the existence of hidden multistability but also underscore the rich and intricate interplay among drive, dissipation, and nonlinearity.

*Theoretical model with multistability and hidden states.* As sketched in Fig. 1(a), we consider a nonlinear driven-dissipative double-oscillator model, where the oscillators, each with intrinsic loss $\gamma_0$, are coupled through reciprocal coupling $g$. The nonlinearity arises from the self-defocusing and self-focusing Kerr effects [32-35] of oscillators 1 and 2, respectively. It is characterized by the intensity-dependent resonant frequencies $\omega_1 = -\Delta\omega + k_1|\psi_1|^2$ and $\omega_2 = +\Delta\omega - k_2|\psi_2|^2$, where $\pm\Delta\omega$ are linear frequencies, $k_{1,2} > 0$ are nonlinear coefficients and $\psi_{1,2}$ represent state components. A harmonic drive $D = A_d e^{-i\omega_d t}$ is applied to oscillator 2, with $A_d$ and $\omega_d$ being the driving amplitude and frequency, respectively. The dynamics of this model is governed by the Gross-Pitaevskii equation

$$i\frac{d\Psi}{dt} = \begin{bmatrix} \omega_1 - i\gamma_0 & g \\ g & \omega_2 - i\gamma_0 \end{bmatrix}\Psi + \begin{bmatrix} 0 \\ 1 \end{bmatrix}D, \quad (1)$$

where $\Psi = e^{-i\omega_d t}[\psi_1, \psi_2]^T$. To avoid strong nonlinear phenomena like limit cycles and chaos [36,37], we focus on the weakly nonlinear regime, where the static solutions take the form $\psi_{1,2} = A_{1,2}e^{-i\theta_{1,2}}$, with $A_{1,2}$ and $\theta_{1,2}$ representing the magnitudes and phases, respectively. More concretely, from Eq. (1) we obtain a higher-order equation for the intensity $I_1 = A_1^2$ of oscillator 1

$$[g^2 - 2(\Lambda_2 - k_2\Pi g^{-2}I_1)(\Lambda_1 + k_1 I_1) + 2\gamma_0^2 + \Pi g^{-2}\gamma_0^2 \\ +\Pi g^{-2}(\Lambda_2 - k_2\Pi g^{-2}I_1)^2] I_1 - A_d^2 = 0, \quad (2)$$

where $\Pi = (-\Delta\omega - \omega_d + k_1 I_1)^2 + \gamma_0^2$, $\Lambda_1 = -\Delta\omega - \omega_d$, and $\Lambda_2 = \Delta\omega - \omega_d$. Using the Routh-Hurwitz criterion, the meaningful solutions for $I_1 > 0$ can be classified into stable and unstable states of the system (see details in Supplemental Material).

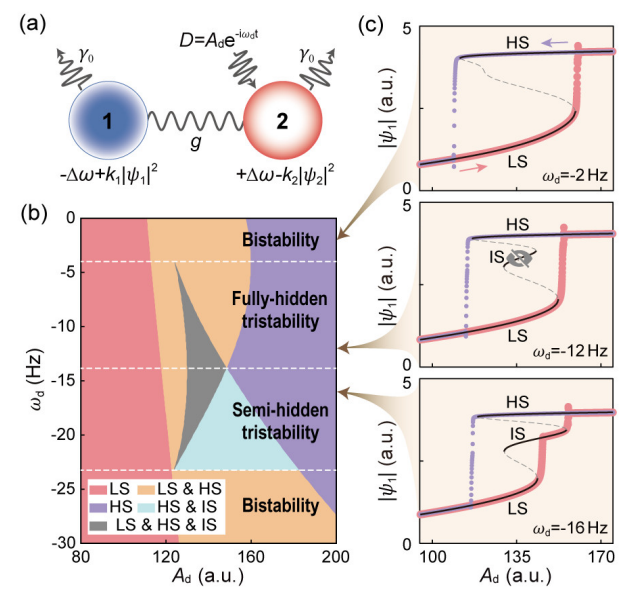

FIG. 1. Theoretical model. (a) Nonlinear driven-dissipative double-oscillator model. Here $\omega_1 = -\Delta\omega + k_1|\psi_1|^2$ and $\omega_2 = \Delta\omega - k_2|\psi_2|^2$ are intensity-dependent resonant frequencies of the two oscillators, $\gamma_0$ is intrinsic loss, $g$ is the coupling between the oscillators, and $D = A_d e^{-i\omega_d t}$ characterizes the drive applied to oscillator 2. (b) Phase diagram. The model supports three stable states, dubbed lower state (LS), higher state (HS), and intermediate state (IS), under different driving amplitude $A_d$ and frequency $\omega_d$. According to the overlap of states at a fixed $\omega_d$, the model exhibits bistability, fully-hidden tristability, and semi-hidden tristability. (c) State evolutions under adiabatic sweeps of $A_d$, corresponding to bistability (top), fully-hidden tristability (middle), and semi-hidden tristability (bottom). The black solid and gray dashed lines specify the stable and unstable states, respectively. The red and purple dots represent the numerical stable states during the forward and backward sweeps of $A_d$, respectively. The parameters used for the calculations are: $\Delta\omega = 43$ Hz, $\gamma_0 = 13$ Hz, $g = 17.5$ Hz, $k_1 = 3.8$, and $k_2 = 1$.

To facilitate our subsequent experiments, the model parameters are set to $\Delta\omega = 43$ Hz, $\gamma_0 = 13$ Hz, $g = 17.5$ Hz, $k_1 = 3.8$, and $k_2 = 1$. Figure 1(b) displays the phase diagram of stable states for $A_d \in [80, 200]$ and $\omega_d \in [-30, 0]$ Hz. Due to the complex interplay of drive, loss, and nonlinearity, the model can exhibit two or three coexisting stable states, depending on the driving parameters $A_d$ and $\omega_d$. Based on the magnitude of $\psi_1$, the stable states are labeled as lower state (LS), higher state (HS), and intermediate state (IS). Meanwhile, according to the overlap of stable states at a fixed $\omega_d$, the phase diagram is divided into different regimes. These include bistability, where only LS and HS coexist within a range of $A_d$; fully-hidden tristability, where an additional IS emerges within the LS-HS overlap range; and semi-hidden tristability, where IS partially extends beyond the overlap. Figure 1(c) exemplifies the three multistabilities under adiabatic sweeps of $A_d$. In the case of $\omega_d = -2$ Hz, as $A_d$ increases, the system evolves along the LS path to a bifurcation point and then jumps up to HS; conversely, as $A_d$ decreases, it follows the HS path to another bifurcation point and finally jumps down to LS. The forward and backward sweeps together form a characteristic hysteresis loop of bistability. Surprisingly, in the case of fully-hidden tristability ($\omega_d = -12$ Hz), despite the presence of additional IS, the system exhibits similar hysteresis that renders IS completely hidden. In semi-hidden tristability ($\omega_d = -16$ Hz), the system transitions from LS to IS and subsequently to HS during a forward $A_d$ sweep, in contrast to the direct transition from HS to LS under a backward sweep. This leads to a partially observable IS, while the remainder is hidden within the LS-HS overlap.

Here, we emphasize that multistability in nonlinear systems is typically highly sensitive to both intrinsic and external parameters. As exemplified by our model [Fig. 1(b)], even a slight shift in the driving frequency can steer the system into entirely different multistable regimes. This pronounced sensitivity carries two crucial implications: on the one hand, it may result in unanticipated outcomes in laboratory experiments or real-world applications; on the other hand, it imposes stringent demands on experimental setups—necessitating precise control over system parameters and accurate measurements to reliably identify hidden multistability.

*Experimental implementation of a nonlinear acoustic platform.* To realize the theoretical model introduced above, we construct an acoustic coupled binary-cavity system with balanced self-focusing and -defocusing Kerr nonlinearities. Remarkably, while acoustic coupled-cavity systems have been widely employed in exploring (linear) topological and non-Hermitian physics [38-45], those incorporating nonlinear effects have yet to be realized. As depicted in Fig. 2(a), the air-filled cavities 1 and 2 feature intrinsic dipole resonant frequencies of 3954 Hz and 4040 Hz, respectively, yielding a central reference frequency $\omega_o$ = 3997 Hz and the desired $\Delta\omega = 43$ Hz for the system. Reciprocal acoustic coupling, $g = 17.5$ Hz, is generated via two narrow tubes connecting the cavities. The cavity losses $\gamma_0$, originally 35 Hz, are compensated to 13 Hz using two linear electroacoustic self-gain modules [42-45], each consisting of a microphone, a preamplifier, a phase shifter, and a speaker [Fig. 2(a), top]. (Notably, the phase shifter provides additional flexibility to compensate for the phase shift introduced by the circuits [45], thereby maintaining the original resonant frequency of the cavity.) Unlike the Kerr-type nonlinearities inherent in optical media or cavity magnonic systems [46-50], airborne sound waves generally exhibit weak nonlinearities that cannot be readily harnessed. Here, we realize acoustic Kerr nonlinearities by ingeniously introducing two nonlinear electroacoustic feedback modules [Fig. 2(a), bottom]. In each module, the sound signal within a cavity is collected by a microphone, modulated by a preamplifier, voltage-controlled amplifier and phase shifter, and then fed back into the cavity via a speaker. Meanwhile, the gain factor of the voltage-controlled amplifier is dynamically adjusted by a microcomputer, which is custom-coded to respond to the pre-amplified signal (proportional to sound pressure in the cavity). Ultimately, our electroacoustic feedback-assisted experimental platform offers exceptional programmability, providing precise parameter control for probing and manipulating complex nonlinear dynamics.

With nonlinear modules individually applied to cavities 1 and 2, we measured the transmission spectra $|S_{12}|$—the response amplitude of cavity 1 to a source in cavity 2—under varying sound intensities $|\psi_1|^2$ and $|\psi_2|^2$. By numerically fitting the spectra [Fig. 2(b)] with the established coupled-mode theory [51,52], we extracted the resonant frequencies of the two cavities [Fig. 2(c)]. As required by our preset parameters ($\Delta\omega = 43$ Hz, $k_1 = 3.8$, and $k_2 = 1$), the data follow the quantitative relations $\omega_1 \approx \omega_o + (-43 + 3.8l|\psi_1|^2)$ Hz and $\omega_2 \approx \omega_o + (43 - l|\psi_2|^2)$ Hz. (Note that the shift of central frequency to $\omega_o$ and the scaling factor $l = 12$ do not affect the essential physics of the system). Specifically, when the nonlinear module in cavity 1 operates independently, $\omega_1$ exhibits a blueshift with increasing $|\psi_1|^2$, indicating the desired self-defocusing nonlinearity in cavity 1. In contrast, when the nonlinear module in cavity 2 is active, $\omega_2$ undergoes a redshift as $|\psi_2|^2$ increases, suggesting the expected self-focusing nonlinearity in cavity 2. Notably, although the absolute frequency shift is measured on a Hz scale, its relative value with respect to the operating frequency ($\omega_o$) reflects a controllable nonlinear strength comparable to that observed in other systems, such as the magnonic system discussed in Ref. 5. More experimental details are provided in Supplemental Material.

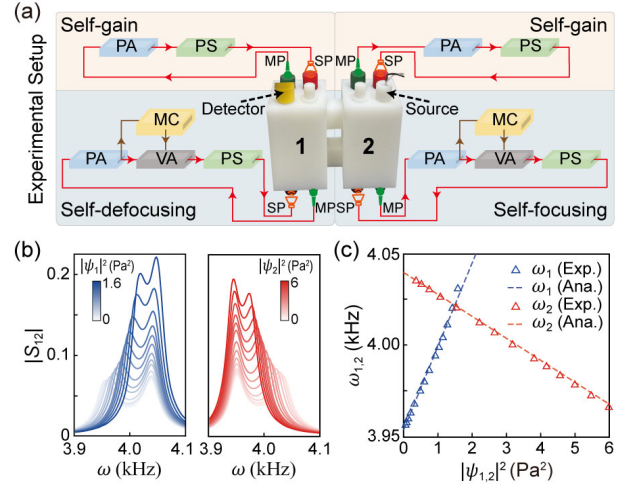

FIG. 2. Nonlinear acoustic coupled binary-cavity system. (a) Experimental setup. Two air-filled acoustic cavities act as oscillators, with narrow tubes between them producing reciprocal coupling. To tune cavity losses, two linear electroacoustic self-gain modules (top) are employed, each comprising microphone (MP), preamplifier (PA), phase shifter (PS), and speaker (SP). To generate self-defocusing and self-focusing nonlinearities, two nonlinear electroacoustic feedback modules (bottom) are introduced, each consisting of MP, PA, voltage-controlled amplifier (VA), PS, microcomputer (MC) and SP. (b) Sound intensity-dependent transmission spectra $|S_{12}|$, in which the nonlinear modules are individually implemented to cavities 1 (left) and 2 (right), through which one can extract their resonant frequencies. (c) Quantitative relations between the resonant frequencies and sound intensities. The linearly-shaped blueshift of $\omega_1$ with $|\psi_1|^2$ and redshift of $\omega_2$ with $|\psi_2|^2$ indicate the presence of self-defocusing and self-focusing Kerr nonlinearities in cavities 1 and 2, respectively.

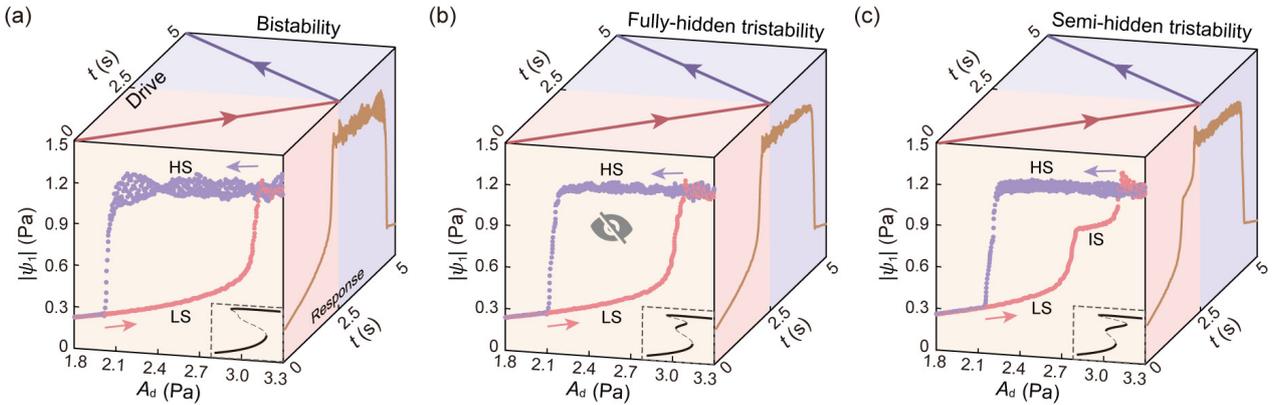

FIG. 3. Experimental characterization of hysteresis loops under continuous sound sweeping. (a)-(c) Hysteresis loops measured for bistability, fully-hidden tristability, and semi-hidden tristability, where the driving frequencies are 3995 Hz, 3985 Hz, and 3981 Hz, respectively. In each cube, the top surface illustrates the temporal variation of driving sound pressure in cavity 2, the side surface displays the sound pressure response in cavity 1, and the front surface shows the state evolutions under sound sweeping. Insets: theoretically predicted multistabilities. In comparison to the partially observed IS in semi-hidden tristability, the IS cannot be detected through sound sweeping in fully-hidden tristability.

*Experimental evidence of multistability and hidden states.* We first demonstrate the hysteresis loops of multistabilities in our system through continuous sound pressure sweeping. To achieve the bistability predicted in Fig. 1(c), we drive the system from cavity 2 at a monochromatic sound of 3995 Hz (equivalent to $\omega_d = -2$ Hz in our theoretical model, with the central reference frequency $\omega_o = 3997$ Hz). The pressure amplitude $A_d$ is continuously increased and decreased during the forward and backward sweeps, each of which lasts 2.5 s [Fig. 3(a), top surface]. (For experimental feasibility, we choose a total measurement duration of $t = 5s$, allowing us to reliably track the state evolution without compromising adiabaticity). To monitor the state evolution, we measure the sound pressure response $|\psi_1|$ in cavity 1 [Fig. 3(a), side surface]. By combining the drive-response data, we observe a hysteresis loop of bistability in the $A_d$-$|\psi_1|$ plane [Fig. 3(a), front surface]. During the forward sweep, as $A_d$ linearly increases from 1.80 Pa to 3.30 Pa, the system evolves gradually along the LS and undergoes a sharp transition to the HS at $A_d \approx 3.05$ Pa, after which the system stabilizes at the HS. Conversely, during the backward sweep, the system remains in the HS until $A_d$ decreases to 2.10 Pa, at which it abruptly switches back to the LS. Remarkably, when the driving frequency is tuned to 3985 Hz (equivalently, $\omega_d = -12$ Hz), $|\psi_1|$ traces a characteristic hysteresis loop similar to that of bistability [Fig. 3(b)]. That is, despite the expectation of tristability (see inset), we are unable to reach the theoretically predicted IS in both the forward and backward parameter sweeps. This underscores the nature of the hidden multistability. Further lowering the driving frequency to 3981 Hz (equivalently, $\omega_d = -16$ Hz), the system exhibits a more complex hysteresis structure of multistability [Fig. 3(c)]. During the forward sweep, $|\psi_1|$ undergoes two abrupt jumps at $A_d \approx 2.71$ Pa and 3.07 Pa, transitioning from LS to IS and from IS to HS, respectively. Conversely, in the backward sweep, the system directly evolves from HS to LS, where the predicted hidden states in this semi-hidden tristability, as expected, are not detected again.

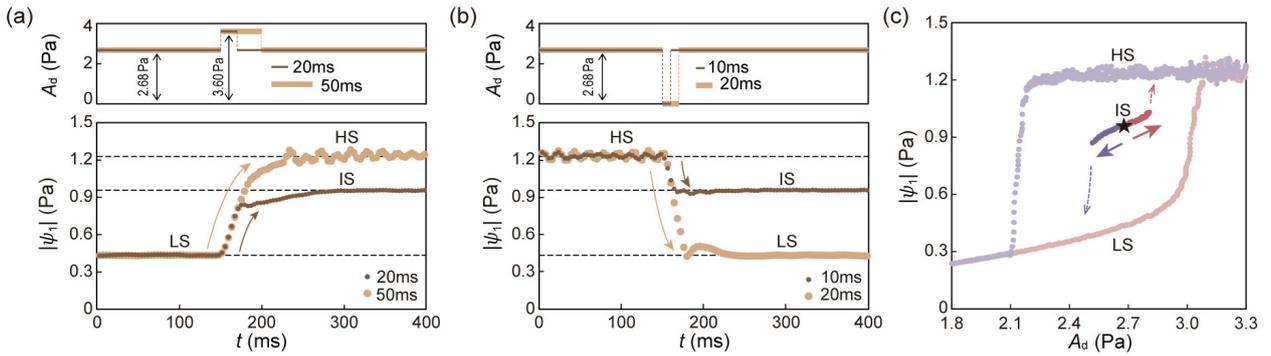

FIG. 4. Experimental observation of hidden states and fully-hidden tristability. (a) Positive sound pulses (top) and the resultant transitions from LS to IS and HS (bottom). A hidden IS is excited from the LS by a positive pulse of 0.92 Pa in height and 20 ms in duration (dark brown), while a HS is converted from the LS by a longer positive pulse of 50 ms duration (light brown). (b) Negative sound pulses (top) and the resultant transitions from HS to IS and LS. As expected, a short negative pulse transitions the HS to a hidden IS, while a long negative pulse switches it to LS. (c) Unveiling of fully-hidden tristability. The complete IS is detected through forward and backward sound sweeps (dark-colored), starting from the hidden IS (star) detected in (a). For comparison, the hysteresis loop from Fig. 3(b) is also displayed (light-colored).

Next, we experimentally reveal the hidden states and fully-hidden tristability by employing an unusual sound pulsing excitation. According to the results in Fig. 3(b), we first drive the system with a sound signal of 3985 Hz and $A_d = 2.68$ Pa to initiate a LS with $|\psi_1| \approx 0.43$ Pa, as shown in Fig. 4(a). At $t = 150$ ms, we apply a positive sound pulse with a height of 0.92 Pa and a duration of 20 ms. This induces a rapid increase in $|\psi_1|$ and pushes the system into the attraction basin of IS. After the driving signal is restored ($t > 170$ ms), $|\psi_1|$ continues to rise and eventually settles at 0.96 Pa —this final value clearly indicate that the system has landed on the hidden IS. Note that both the duration and amplitude of the pulse are crucial for the transition LS → IS. A pulse that is too short (weak) cannot excite the LS into the IS basin, eventually causing the system to return to the LS. Conversely, a pulse that is too long (strong) will drive the system into the attraction basin of HS. For example, as shown in Fig. 4(a), when we extend the pulse duration to 50 ms, the system bypasses the IS basin and directly falls into the HS with $|\psi_1| \approx 1.23$ Pa. (The oscillation of $|\psi_1|$ at the HS can be attributed to the instability of the experimental system under relatively strong nonlinearity). Similarly, we can induce the transitions from HS to the hidden IS and LS through applying negative pulses. As shown in Fig. 4(b), starting from a HS of $|\psi_1| \approx 1.23$ Pa, a negative pulse with a depth of 2.68 Pa and a duration of 10 ms drives the system to the hidden IS, while a longer 20 ms pulse transitions the system directly to the LS. More experimental results can be seen in Supplementary Material. It is worth noting that only a single hidden IS is detected in each sound pulsing excitation. To construct the complete hidden IS path embedded within the hysteresis loop, we further perform forward and backward sound field sweeps, starting from a known hidden IS. As shown in Fig. 4(c), the entire hidden IS within the range of $A_d \in [2.52, 2.81]$ Pa is revealed around the single hidden IS at $A_d = 2.68$ Pa. Combined with the hysteresis loop obtained in Fig. 3(b), now all three stable states of the fully-hidden tristability are clearly identified in our experiments. Similar treatment can be applied to the system of semi-hidden tristability, where the hidden states missed in Fig. 3(c) can be uncovered as expected (see Supplementary Material).

*Conclusions and outlook.* We have experimentally identified the presence of hidden multistability, a subtle and previously unobserved nonlinear phenomenon, using our newly developed nonlinear acoustic platform. Beyond the hysteresis loops uncovered via standard sound field sweeping, the elusive hidden states (and their transitions from known stable states) are explicitly revealed and dynamically controlled via unusual pulse excitation protocol. Our work not only brings the study of hidden multistability to the experimental level, but also inspires innovative control of multistate dynamics in nonlinear systems across a broad spectrum of scientific fields, including optics, mechanics, magnonics, electronic circuits, and even neuroscience and ecology.

Looking ahead, our nonlinear acoustic platform offers exciting opportunities for uncovering even more complex multistable configurations, such as hierarchical hidden states, through the introduction of additional coupled cavities or drives. Its high programmability, powered by electroacoustic feedback modules, permits precise and flexible control over onsite frequencies, gain/loss distributions, and reciprocal/nonreciprocal couplings as functions of sound intensity, time, and other parameters. The resultant scalability facilitates the further integration of diverse advanced physical effects—including non-Hermiticity, synthetic gauge fields, band topology, and spatiotemporal modulations—making our experimental platform a versatile tool for exploring their rich interplay with nonlinear dynamics. Owing to the universal nature of these physical effects across different systems, our work not only establishes a foundation for developing intelligent multifunctional acoustic metamaterials, but also paves the way for exploring a wide spectrum of cutting-edge physical phenomena beyond acoustics, including nonlinear bulk-edge correspondence [53-57], nonlinear topological solitons [47,48,58-60], nonlinear Thouless pumping [61-64], nonlinear skin effects [35,65-69], as well as nonlinear exceptional points [70,71].

**Acknowledgements**
This work was supported by the National Natural Science Foundation of China (Grants No. 12374418, No. 12104346, and No. 12304495), the Natural Science Foundation of Hubei Province of China (No. 2024AFB654), and the Fundamental Research Funds for the Central Universities, and the National Key R&D Program of China (Grant No. 2023YFA1406900).